\documentclass[aps,prl,preprint,groupedaddress]{revtex4-1}

\usepackage{amsmath}
\usepackage{wrapfig, blindtext, textcomp, epsfig}
\usepackage[font={scriptsize}]{caption}
\usepackage{caption}
\usepackage{subcaption}
\usepackage{float}
\usepackage{graphicx}
\usepackage{pgf}

\newcommand{\pt}{\ensuremath{p_{\mathrm{T}}}}
\newcommand{\piz}{\ensuremath{\pi^{0}}}
\newcommand{\PbPb}{\ensuremath{\mbox{Pb--Pb}}}
\newcommand{\pp}{\ensuremath{\mathrm {p\kern-0.05em p}}}

\newcommand{\sqrtS}{\ensuremath{\sqrt{s}}}

\newcommand{\MeVc}{\ensuremath{\mathrm{MeV}\kern-0.05em/\kern-0.02em c}}
\newcommand{\GeVc}{\ensuremath{\mathrm{GeV}\kern-0.05em/\kern-0.02em c}}
\newcommand{\MeV}{\ensuremath{\mathrm{MeV}}}

\begin{document}

\title{Photon and neutral pion production \\ in pp and 
	Pb-Pb collisions at LHC energies \\ in the ALICE experiment}

\author{Podist Kurashvili for the ALICE Collaboration}

\affiliation{National Centre for Nuclear Research
               Warsaw, Poland}

\date{\today}


\begin{abstract}
This talk is dedicated to the measurements of direct photons and neutral 
pions in pp at $\sqrt{s}$=2.76 and 7 TeV and Pb-Pb collisions at 
$\sqrt{s}$=2.76 TeV per nucleon pair. 
Photons are measured in the ALICE experiment
at mid-rapidity in two electromagnetic 
calorimeters, Photon Spectrometer (PHOS) 
and Electromagnetic Calorimeter (EMCal), and 
via the measurement of the electron-position pairs 
in the Time Projection Chamber (TPC) 
and Inner Tracking System (ITS).
The latter are 
produced in the interaction of photons 
with the material of the detector. 
Neutral pions are identified via the invariant mass analysis of photon pairs. 
The combination of different detection methods allows to measure the 
spectra of particles with high precision over a wide momentum range.
Photons are a unique tool for studying the evolution
of nuclear matter produced in the collision 
as they are emitted
during the different stages of the 
expansion of the initial hot-matter fireball.
Photons interact weakly 
with other particles, escaping almost unchanged and carry information 
about the properties of the system at the space-time point of their emission.
Direct, high-energy photons, formed at the early stage 
of the collision, provide a test of perturbative QCD, set constraints on 
the parton distribution and fragmentation functions 
and determine the initial energy of 
jets produced by high-energy quarks. 
Thermal photons result 
from interactions of particles in both hot QGP 
and cooler Hadron Gas phases,
and carry  information 
about the temperature of the medium. 
The spectra of photons and the effective temperature 
extracted from the thermal photon yield will be presented 
and compared with the results of the PHENIX experiment.
The measurement of neutral pion yields permits a
study of medium-induced energy loss via the transverse
momentum dependence of 
the nuclear modification factor $R_{AA}$,
expressing the modification of \piz \mbox{}
\pt -spectra in Pb-Pb collisions with
respect to \pp \mbox{} interactions,
as a function of collision centrality. 

\end{abstract}


\maketitle

\section{Introduction}
 
 ~
The quark-gluon plasma (QGP) is the
state of matter at high temperature 
characterized by deconfinement
of quarks and gluons \cite{QGP}.
The QGP could have existed in the early hot Universe
up to about 10$^{-6}$ s after
the Big Bang.
After expansion
and cooling, quarks and gluons were confined into
hadrons which we observe in ordinary matter.
Present-day heavy-ion experiments try to
produce extremely hot
matter with vanishing baryon density
by colliding  heavy ions,
which would lead to a temperature above the
quark-gluon plasma transition point $T \sim$ 155 MeV \cite{Martinez}
("little Big Bang").
The hot matter fireball formed in the
collision undergoes subsequent expansion
going through the quark-gluon plasma and hadron gas stages
and ends by freeze-out when the final-state 
particles stream freely out of the reaction zone.
The measurement of direct photons
is one of the most interesting
challenges in exploring the quark-gluon plasma.
Direct photons can be divided in two groups
according to the mechanism of emission.
\emph{Prompt} photons are produced in the initial
parton interactions
and dominate the high-\pt \mbox{} part of the direct photon spectrum.
\emph{Thermal} photons
are formed in secondary
interactions between quarks and gluons
in the hot matter fireball with energy far below that
of original
interactions. 
Photons experience a very weak interaction 
with the matter
and escape almost unperturbed from the reaction zone
carrying information
about the thermal and dynamical properties of
the medium at the point of their emission.
Decay photons, which constitute the major part of
the photon spectrum, help
to reconstruct short-lived particles, such as
neutral pions that have the dominant two-photon decay
channel.
Neutral pions are produced by quark and
gluon fragmentation
in the initial high-energy collisions (hard part)
and from the quark-gluon matter fireball (soft part).
Unlike the direct photon case, their production
is highly affected by the medium.
The study of in-medium modification of the neutral
pion yield and \piz -hadron correlations probes the stopping
power of the hot matter.
Measurement of direct photons in pp and p-Pb
collisions helps to build a reference for studying
the properties of matter in Pb-Pb collisions.
These measurements also provide a tool to probe the perturbative
regime of Quantum Chromodynamics and
and tests the validity of the initial parton
distributions and fragmentation functions used in current particle
generators.”
In this talk we present results obtained in the ALICE LHC experiment
during the Run 1 in 2010-2011 for proton-proton and lead-lead interactions
at 0.9, 2.76 and 7 TeV.
%


\section{Photon measurements in ALICE}

\begin{wrapfigure}{R}{0.45 \textwidth}
				\vspace{-5pt}
				\begin{center}
				\includegraphics[width=50 mm]{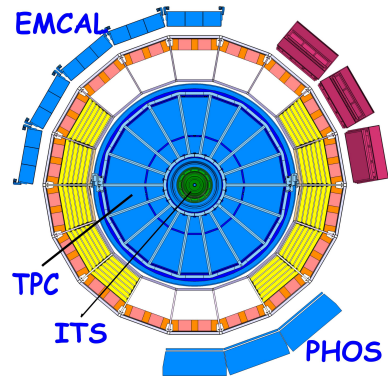}
			\end{center}
			\caption{\label{fig:CrossSec}ALICE detectors used in detection of photons.}
			\end{wrapfigure}

~
The ALICE experiment is a dedicated
heavy-ion experiment designed to study the quark-gluon plasma
and one of four major experiments 
at the Large Hadron Collider (LHC)
at CERN.
The ALICE detector system is designed to cope with the high multiplicities
of final state particles in the mid-rapidity region
and combines different techniques of measurement for
different kinds of particles.
The ALICE experiment is the only LHC experiment 
with the capability
to measure and identify particles
with low $\pt$  \mbox{} ($\leq$ 1 GeV/$c$). 
The full description of the experimental setup and techniques
used in measurements can be found in \cite{ALICE}.
Below we shall concentrate on photon detection
with detectors highlighted in Figure \ref{fig:CrossSec}.

There are two methods
of photon measurements used in the ALICE experiment.
The first, called photon conversion method (PCM), is based
on detecting photons by their conversions to 
electron-positron pairs 
in the material of the detectors.
Electron-positron pairs are reconstructed in the
tracking detectors, ITS consisting
of six layers of silicon detectors  and
gaseous
TPC.
The experimental setup 
provides high acceptance for this method
 (full azimuthal angle,
$|\eta|<$0.9), although the probability of the photon conversion 
is rather low ($\sim$8.5\%)
due to the small material budget of the detectors.
The PCM is characterized by high precision 
of energy measurement at low transverse momenta,
$\pt \leq$ 1 GeV/$c$.
Another method is based on calorimetry. 
Two calorimeters,
PHOS and EMCal, are designed for high precision photon measurements
over a wide dynamical range.
The three PHOS modules consist of 10752
PBWO$_4$ crystals with 2.2$\times$2.2 cm$^2$
cross section and 18 cm depth, corresponding to
approximately 20 radiation lengths.
The acceptance of PHOS is $|\eta|<$0.12 and 60 degrees of
azimuthal angle.
Its high granularity and excellent energy resolution
allow the detection of photon
pairs from \piz \mbox{} decays at high \pt .
Good time resolution makes possible the application of a
time-of-flight
filter.
The EMCal consists of Pb-scintillator sandwich layers
with cell of transverse size 6$\times$6 cm$^2$ 
and depth 24.6 cm, which also
makes up about 20 radiation lengths.
Its acceptance is $|\eta|<$0.7 and
107 degrees of azimuthal angle.
%


\section{Neutral pions in $\mathrm{\textbf{pp}}$ and \PbPb \mbox{} collisions}

~

Neutral pions are extracted using an invariant mass analysis,
combining all possible pairs of reconstructed photons
and calculating the mass of the pairs:
\begin{equation}
M^2_{inv}=(p_1 + p_2)^2 = 2 \epsilon_1 \epsilon_2 (1- \cos\mbox{} \theta_{12}),
\end{equation}
where, $p_{1,2}$, $\epsilon_{1,2}$ are the four-momenta
and energies of the two photons, respectively and $\theta_{12}$ is 
their opening angle.

\begin{figure}[htb]
\includegraphics[scale=0.7]{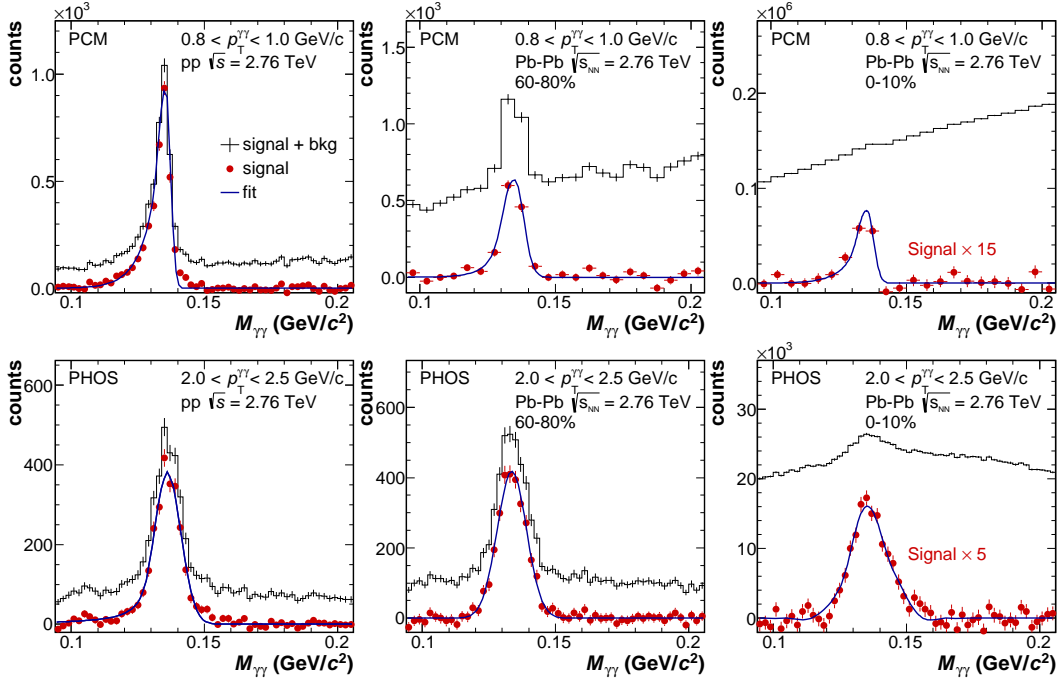}
\caption{\label{fig:InvMass}
Invariant mass spectra in selected \pt\ slices for PCM (upper row) 
and PHOS (lower row) in the $\pi^0$ mass region for \pp\ (left column), 
$60-80\%$ centrality \PbPb\ collisions
(middle column) and $0-10\%$ centrality (right column) \PbPb\ 
collisions \cite{pi0Paper}. 
The histogram and the filled points show the data before and after background 
subtraction, respectively. For the $0-10\%$ class the invariant 
mass distributions after background subtraction were scaled 
by a factor 15 and 5 for PCM and PHOS, respectively, for better 
visibility of the peak.  
The positions and widths of the $\pi^0$  peaks were determined from the fits 
to the invariant mass spectra after background 
subtraction (blue curves).}
\end{figure}

\begin{figure}[H]
\centering
\includegraphics[scale=0.7]{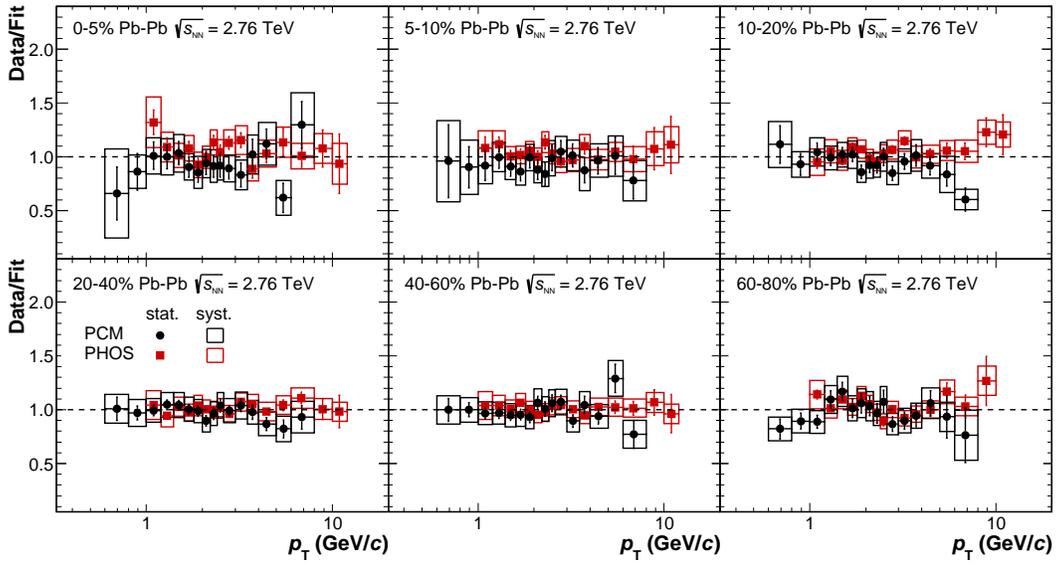}
\caption{Ratio of the fully corrected $\pi^0$ 
spectra in Pb-Pb collisions at $\sqrt{s_{NN}}=2.76$ TeV in six centrality bins measured with PHOS and PCM 
to the fits to the combined result in each bin \cite{pi0Paper}. 
Vertical lines represent statistical uncertainties, the boxes are the systematic uncertainties.}
\label{fig:Comp}
\end{figure}

The resulting $M_{inv}$ distribution is peaked
near $m_{\piz}$=0.135 \GeVc$^2$   
and also contains
combinatorial background, as shown in several
examples
in Figure \ref{fig:InvMass}.
The event mixing technique was used for precise evaluation and subtraction
of the background \cite{pi0Paper}.
As one can see, the extraction method works both in low and high multiplicity
environments.
The raw \piz \mbox{} spectrum obtained from the
invariant mass analysis
is corrected for detector acceptance and efficiency
in order to determine the production cross section.
Figure \ref{fig:Comp} represents the comparison of results obtained
from PCM and calorimetric (PHOS) measurements
in the case of \PbPb\ collisions, where both
histograms are divided by the function fitting the
combined \pt -distribution.
The results
obtained by the two methods agree within errors.

\begin{figure}[H]
\centering
\begin{minipage}{.45\textwidth}
  \centering
  \includegraphics[height=1.6 \textwidth]
  {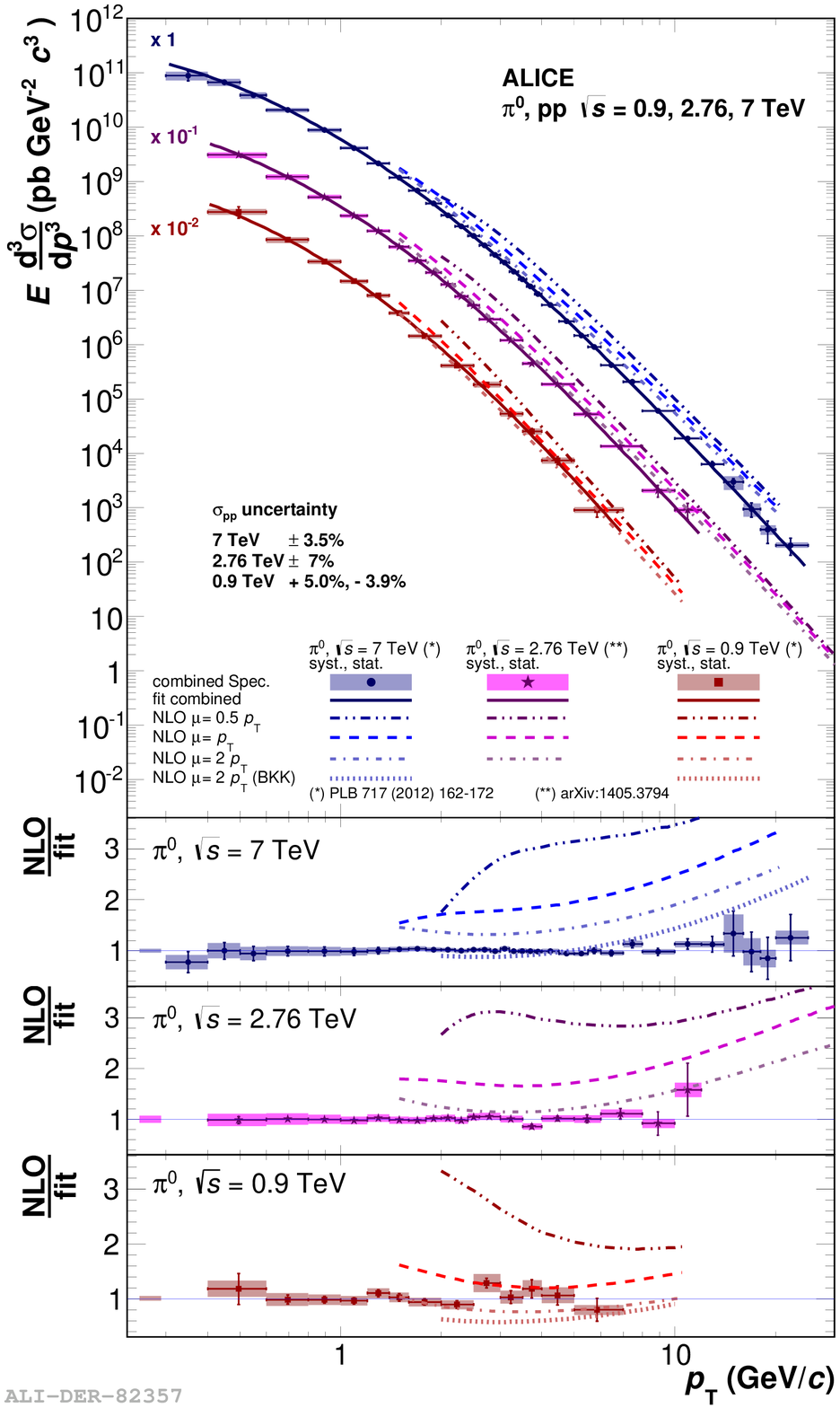}
  \captionof{figure}{Differential cross section of \piz \mbox{} \\
  production in \pp \mbox{} collisions at \sqrtS = 0.9, 2.76 and \\ 
  7 TeV \cite{pi0Paper2}.
  NLO pQCD calculations
  for three \\  factorization
  scales $\mu_F=$ 0.5\pt , \pt , 2\pt \mbox{} \\ are shown.
     }
  \label{fig:pi0pp}
\end{minipage}%
\begin{minipage}{.55\textwidth}
  \centering
  \includegraphics[height=1.31 \textwidth]
  {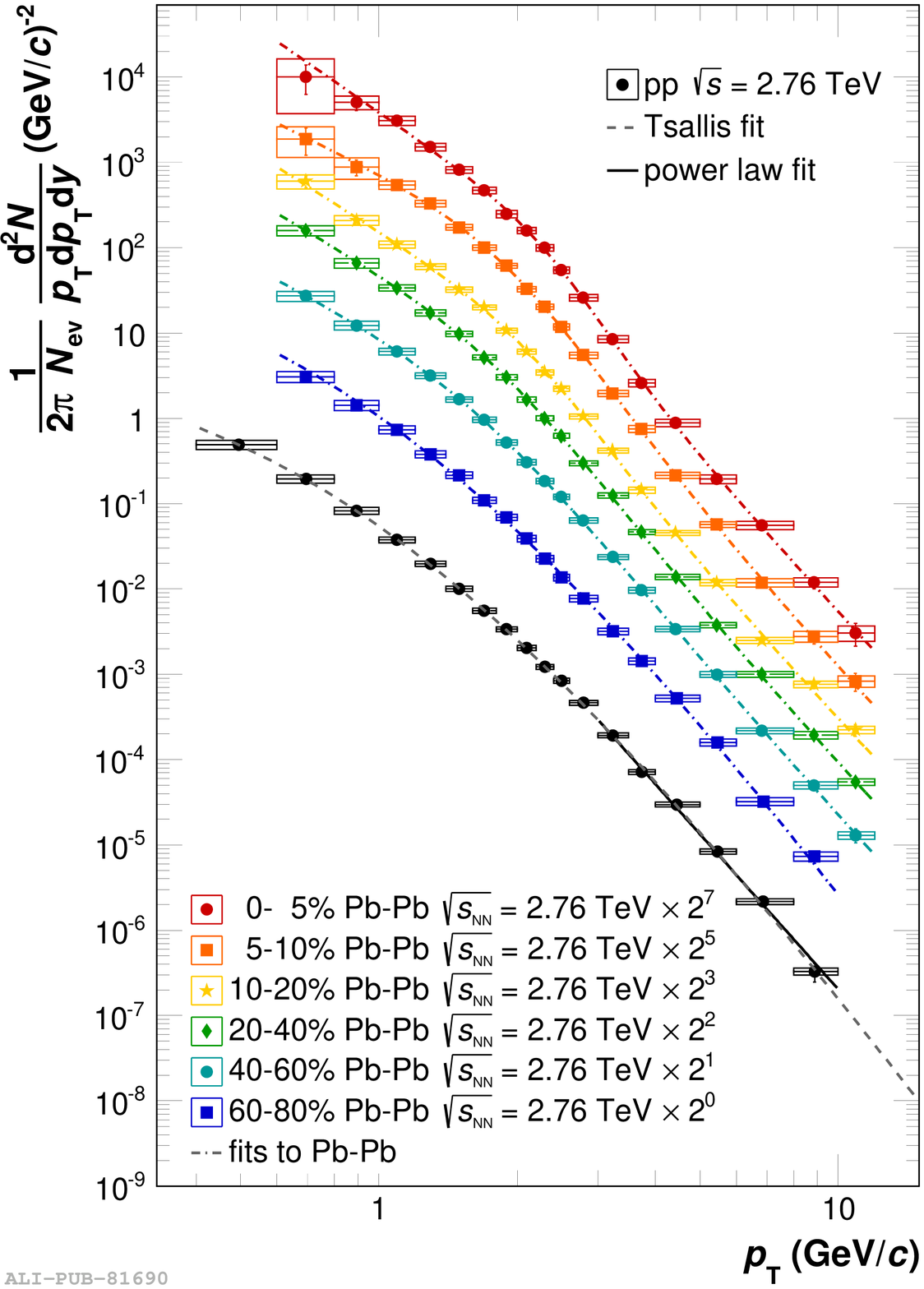}
  \captionof{figure}{Invariant differential yields of neutral
  pions produced in Pb--Pb  and inelastic pp collisions at $\sqrt{s_{NN}}=2.76$ TeV \cite{pi0Paper}. 
 The spectra are the weighted average of the PHOS and the PCM results.}
  \label{fig:pi0PbPb}
\end{minipage}
\end{figure}

The neutral pion production
cross section
is plotted in Figure \ref{fig:pi0pp} for three
different collision energies, 0.9, 2.76 and 7 TeV.
The comparison to Next to Leading Order (NLO)
perturbative QCD calculation
shows a good agreement between theory and experiments
at \sqrtS = 0.9 TeV.
At 2.76 and 7 TeV, theory overestimates the \piz \mbox{}
production
cross section.
This may be connected to the imprecise knowledge of the
gluon fragmentation functions, which are not well defined by
existing data at low \sqrtS \mbox{} \cite{Florian}.

The results for \PbPb \mbox{} invariant yields are shown in
Figure \ref{fig:pi0PbPb}.
Comparison of these results to the \pp \mbox{} yield
gives the nuclear modification factor, $R_{AA}$:
\begin{equation}
R_{AA}(\pt)=\frac{\frac{\mathrm{d}^2 N}{\mathrm{d} \pt \mathrm{d} y}|_{AA}}{\langle T_{AA}\rangle 
\times \frac{\mathrm{d}^2 \sigma}{\mathrm{d}\pt \mathrm{d} y}|_{pp}}.
\end{equation}		

\begin{figure}[H]
\centering
\includegraphics[height=0.4\textwidth]
{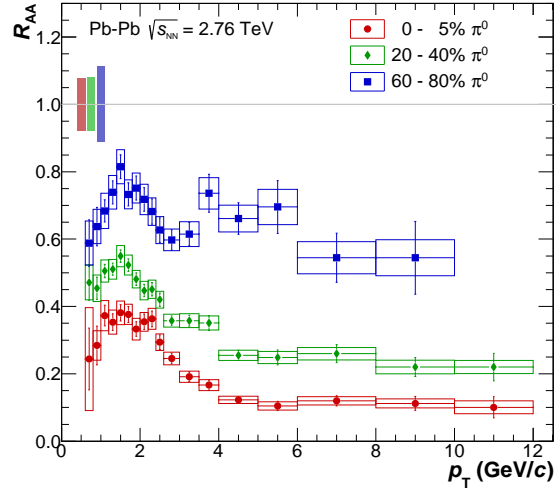}
\caption{ \label{fig:RAA}
 Neutral pion nuclear modification factor $R_{AA}$ for 
 three different centralities (0-5\%, 20-40\%, 60-80\%) in \PbPb \mbox{} collisions 
at $\sqrt{s_{NN}}= 2.76$ TeV \cite{pi0Paper}.
The boxes around unity reflect the uncertainty of the average nuclear 
overlap function $\langle T_{AA}\rangle$ and the normalization uncertainty 
of the pp spectrum added in quadratures.
}
\end{figure}

$R_{AA}$ measures the ratio of the 
\piz\ yield in Pb-Pb collisions compared to the yield in 
pp collisions scaled by the number of binary nucleon-nucleon collisions in a specified centrality range, $N_{coll}$ = $\langle T_{AA} \rangle \sigma^\mathrm{pp}_{inel}$.
A deviation of 
$R_{AA}$ below unity represents a suppression of \piz\ production in \PbPb\ collisions.
Figure \ref{fig:RAA} shows $R_{AA}$ calculated 
for three bins of collision 
centrality.
Pion production is most suppressed in central
\PbPb\ collisions.
The modification factor levels off at $\pt >$4 GeV/$c$
and is equal to 0.1 for most central collisions which is
two times lower than the result obtained by the
PHENIX experiment at RHIC \cite{PHENIX}, which indicates
a larger energy loss compared to RHIC.
Below 4 GeV/$c$ the spectrum is dominated by soft QCD
processes in the medium and boosted by flow, 
resulting in growth of $R_{AA}$ at small \pt .
  

\newpage

\section{Neutral pion angular correlations}

\begin{wrapfigure}{R}{0.6 \textwidth}
				\vspace{-20pt}
				\begin{center}
				\includegraphics[scale=0.5]{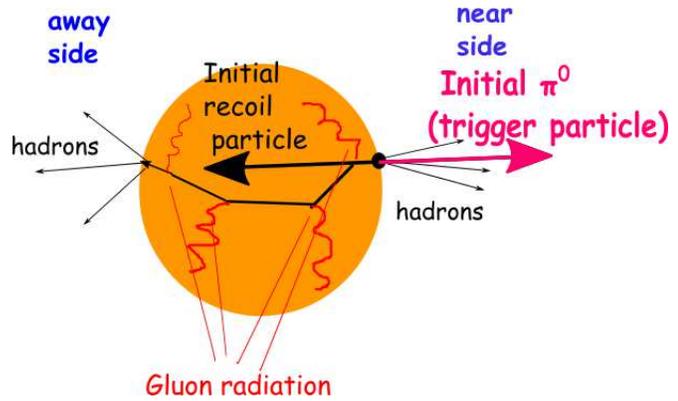}
			\end{center}
			\caption{\label{fig:jets}
			Graphical representation 
			of trigger \piz \mbox{}, associated hadrons and
			near and away sides.}
			\end{wrapfigure}

~
High transverse momentum quarks and gluons produced in the
collisions hadronize creating a spray of hadrons emitted
in narrow cones around the direction of the scattered partons
called jets.
A jet is commonly accompanied by another parton
flying in opposite direction.
Two opposite jets produced in quark-gluon plasma should
undergo different degree of suppression if their paths in medium
are different.
In \piz -hadron correlation measurements,
pions with high \pt\ are presumed to 
be the 
leading particle of hadron jets
produced near the surface of quark-gluon plasma.
In that case, the jet of the corresponding recoil particle
should be observed in the opposite direction as is shown
in Figure \ref{fig:jets}.

\begin{figure}
\centering
\begin{subfigure}{.5\textwidth}
  \centering
  \includegraphics[width=\textwidth]
{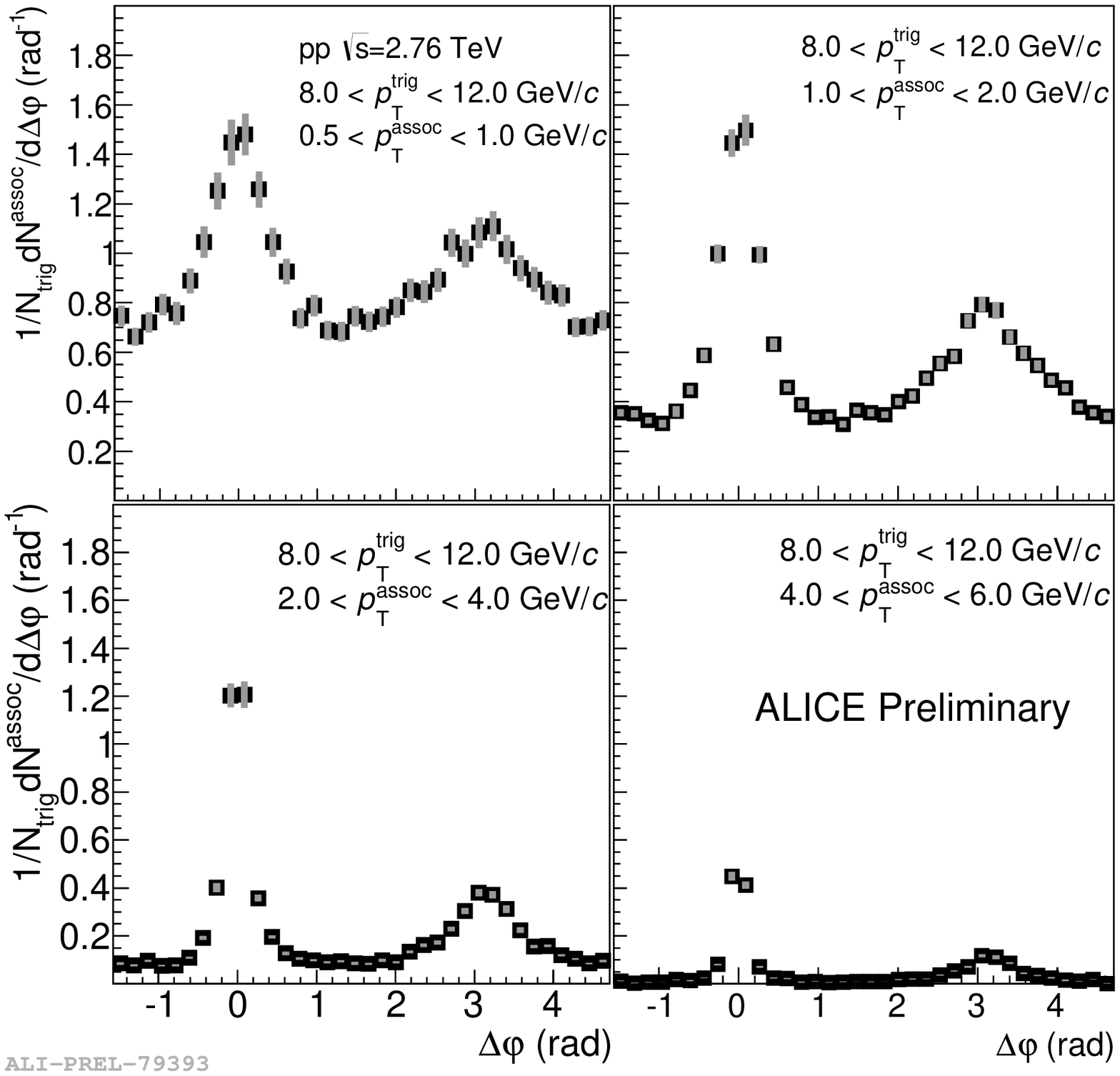}   
\end{subfigure}%
\begin{subfigure}{.5\textwidth}
  \centering
  \includegraphics[width=\textwidth]
  {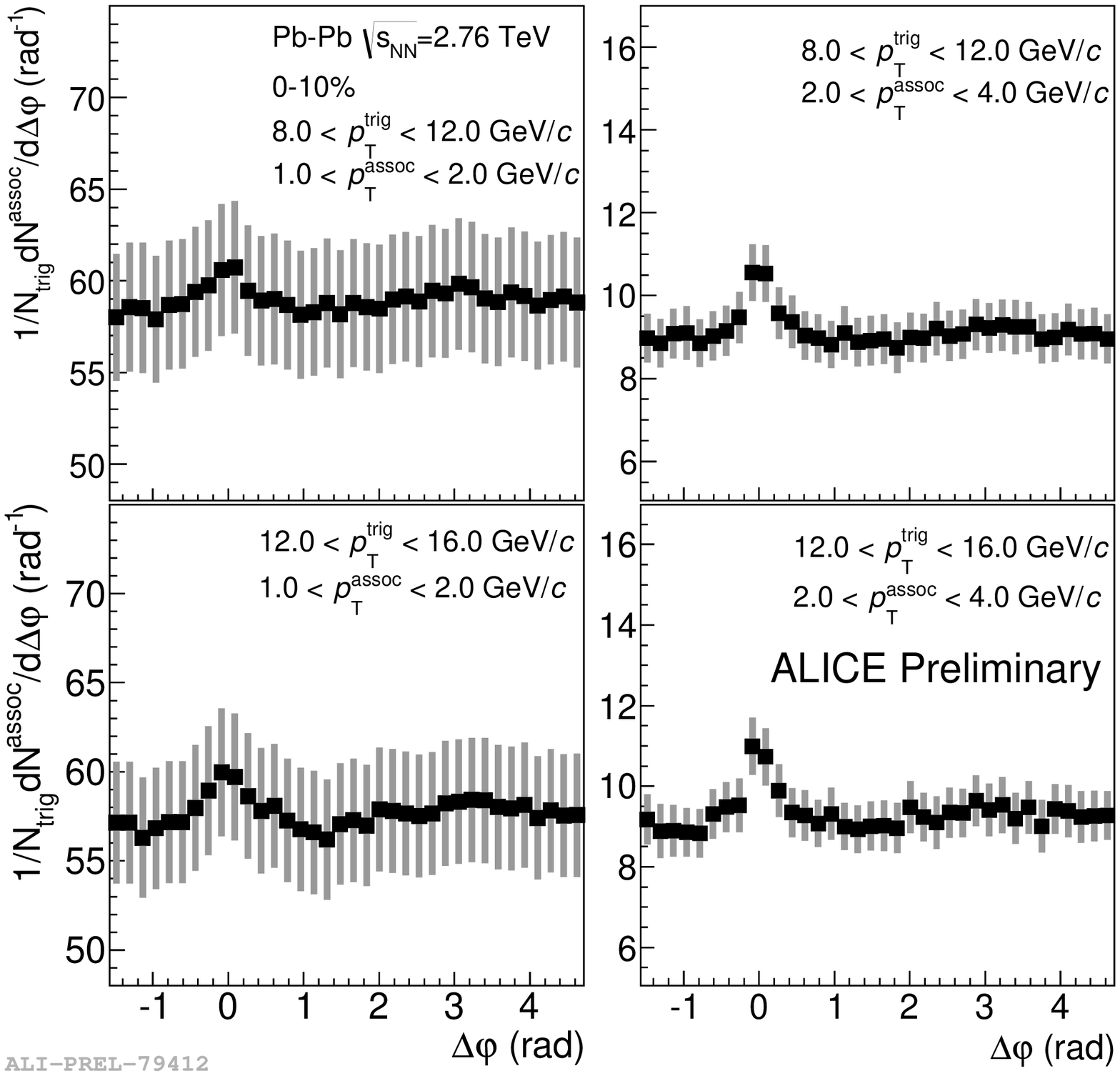}
\end{subfigure}
\caption{
Azimuthal angle distribution of 
\piz -hadron correlations 
for different bins of transverse momentum of associated
particles
with the trigger \piz\
in \pt\ bins 
8-12 GeV/$c$ for pp collisions at \sqrtS =2.76 TeV  (left) and
8-12 GeV/$c$ and 12-16 GeV/$c$ for \PbPb\
collisions at $\sqrt{s_{NN}}=$2.76 TeV, centrality 0-10\% (right).
}
\label{fig:Corr}
\end{figure}

We looked for correlations between these pions, which were taken
as trigger particles, and charged hadrons (associated particles)
by studying the azimuthal angle
$\varDelta \varphi = \varphi^{trig}-\varphi^{assoc}$.
Charged hadrons were detected by the
tracking detectors and
pions were measured with EMCAL using a
combination of invariant mass and cluster shape analyses.
The latter is based on the difference of the second moment
of the cell energy distribution in the cluster.
The greater eigenvalue of this tensor
significantly varies
depending on whether a 
cluster was produced by a single photon or
by a closely separated pair of photons from \piz \mbox{}  decay.
Defining the charged hadron yield per
trigger particle,
$Y^{\mathrm{\textit{pp}}}(\pt^{\pi^0}, \pt^{h^{\pm}})$
and
$Y^\mathrm{\textit{AA}}(\pt^{\pi^0}, \pt^{h^{\pm}})$,
for \pp\ and \PbPb\ collisions, respectively,
the observable that quantifies medium effects is the
modification factor:

\begin{equation}
I_\mathrm{\textit{AA}}(\pt^{\pi^0}, \pt^{h^{\pm}})= 
			\frac{Y^\mathrm{\textit{AA}}(\pt^{\pi^0}, \pt^{h^{\pm}})}
			{Y^{\mathrm{\textit{pp}}}(\pt^{\pi^0}, \pt^{h^{\pm}})}
\end{equation}
The results for correlations for \pp \mbox{} and \PbPb \mbox{}
collisions are presented in Figure \ref{fig:Corr}.
As one can see, there are two maxima close 
to $\varDelta \varphi=$ 0 (near side)
and $\varDelta \varphi= \pi$ (away side) in \pp \mbox{}
collisions.
%
%

\begin{figure}[H]
\begin{center}
\includegraphics[scale=0.35]{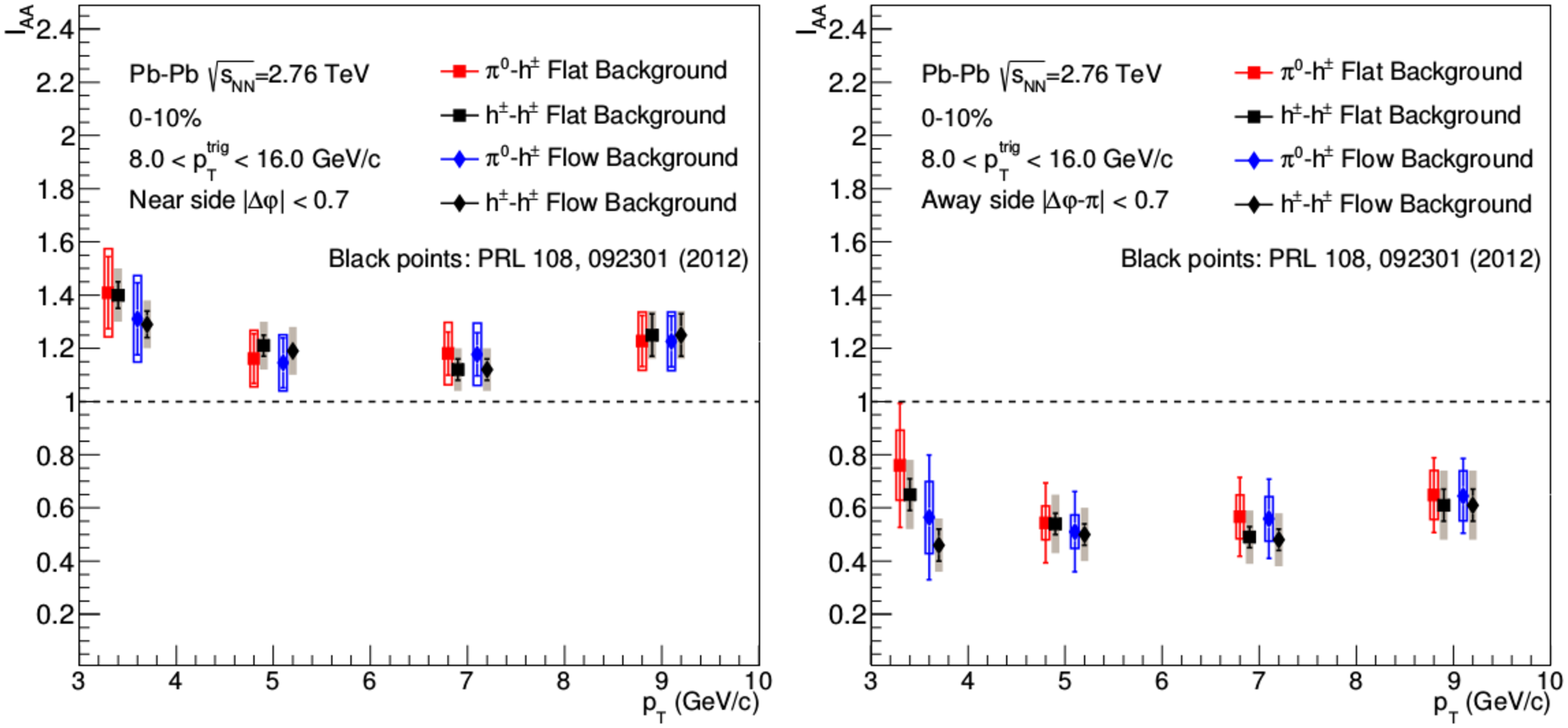}
\end{center}
\caption{\label{fig:IAA}
Modification of per-trigger yield of charged hadrons, 
$I_{AA}$, on the near (left) and away (right)
side for \piz -hadron correlations 
at 0-10\% in \PbPb \mbox{} collisions at
$\sqrt{s_{NN}}=$2.76 TeV.
}
\end{figure}

%
The modification factor at near and away sides is presented in
Figure \ref{fig:IAA}.
The away-side correlation exhibits suppression by a factor of
two due to energy loss in the medium.
The near-side enhancement by about 20\%
can be explained by several factors,
such as a change of the fragmentation function, different
fractions of quark and gluon jets in the
medium compared to the \pp\ case, and bias of parton \pt\
distribution due to in-medium energy loss \cite{Enhance}.

\newpage

\section{Direct photons}
		
\begin{figure}[H]
\centering
\begin{subfigure}{.5\textwidth}
  \centering
  \includegraphics[width=\textwidth]{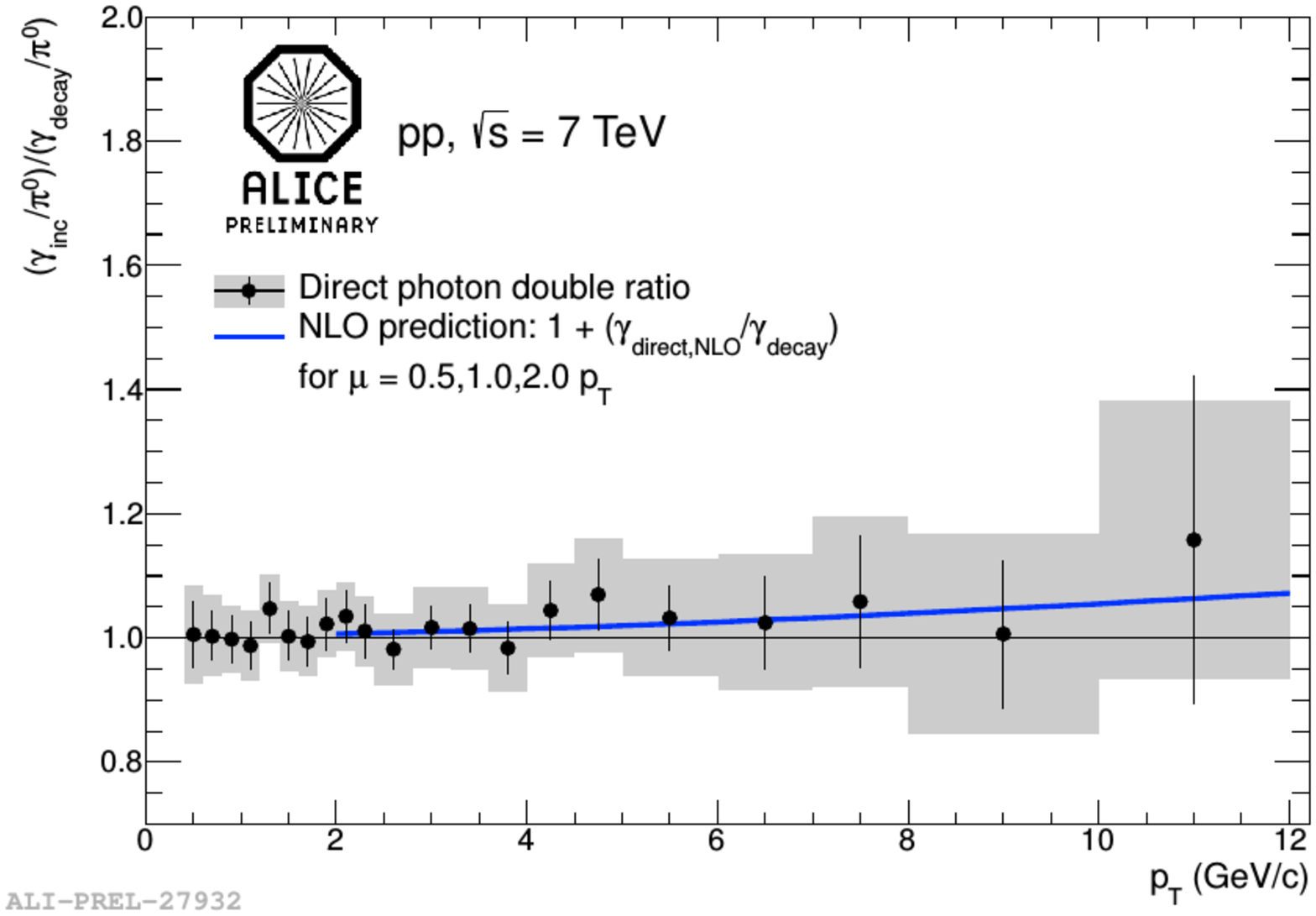}   
\end{subfigure}%
\begin{subfigure}{.5\textwidth}
  \centering
  \includegraphics[width=\textwidth]
  {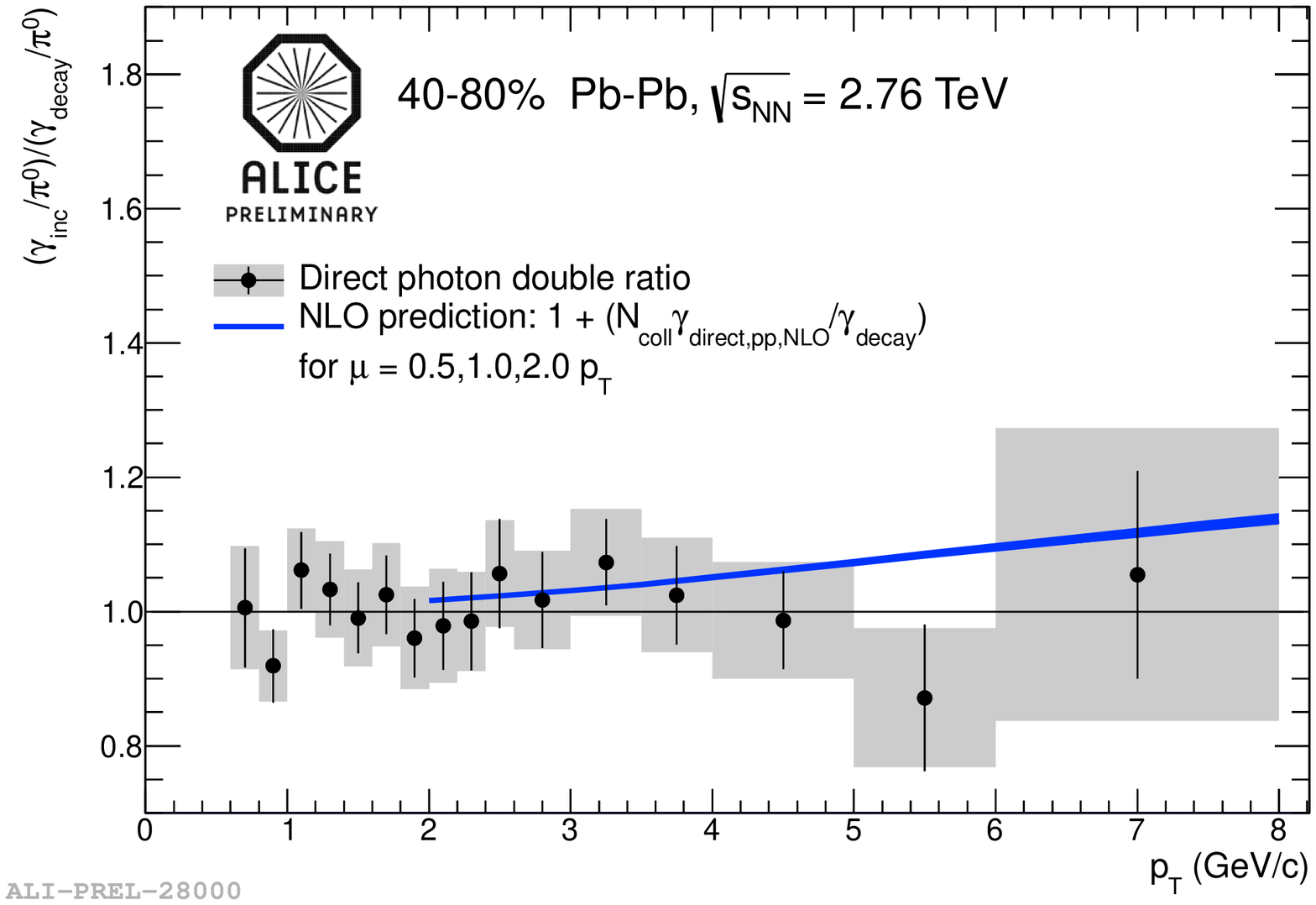}
\end{subfigure}
\caption{
Direct photon double ratio and NLO pQCD prediction (blue curve) 
in \pp \mbox{} collisions at \sqrtS = 7 TeV (left)
and
   at 40-80\% 
  in \PbPb \mbox{} collisions at $\sqrt{s_{NN}}=2.76$ TeV (right)
  \cite{Wilde}. 
  The spread of the blue curve corresponds to different factorization
  scales $\mu_F=$0.5\pt , \pt, 2\pt \mbox{}.
}
\label{fig:Double_pp}
\end{figure}

\begin{figure}[H]
\centering
\begin{subfigure}{.5\textwidth}
  \centering
  \includegraphics[width=\textwidth]
  {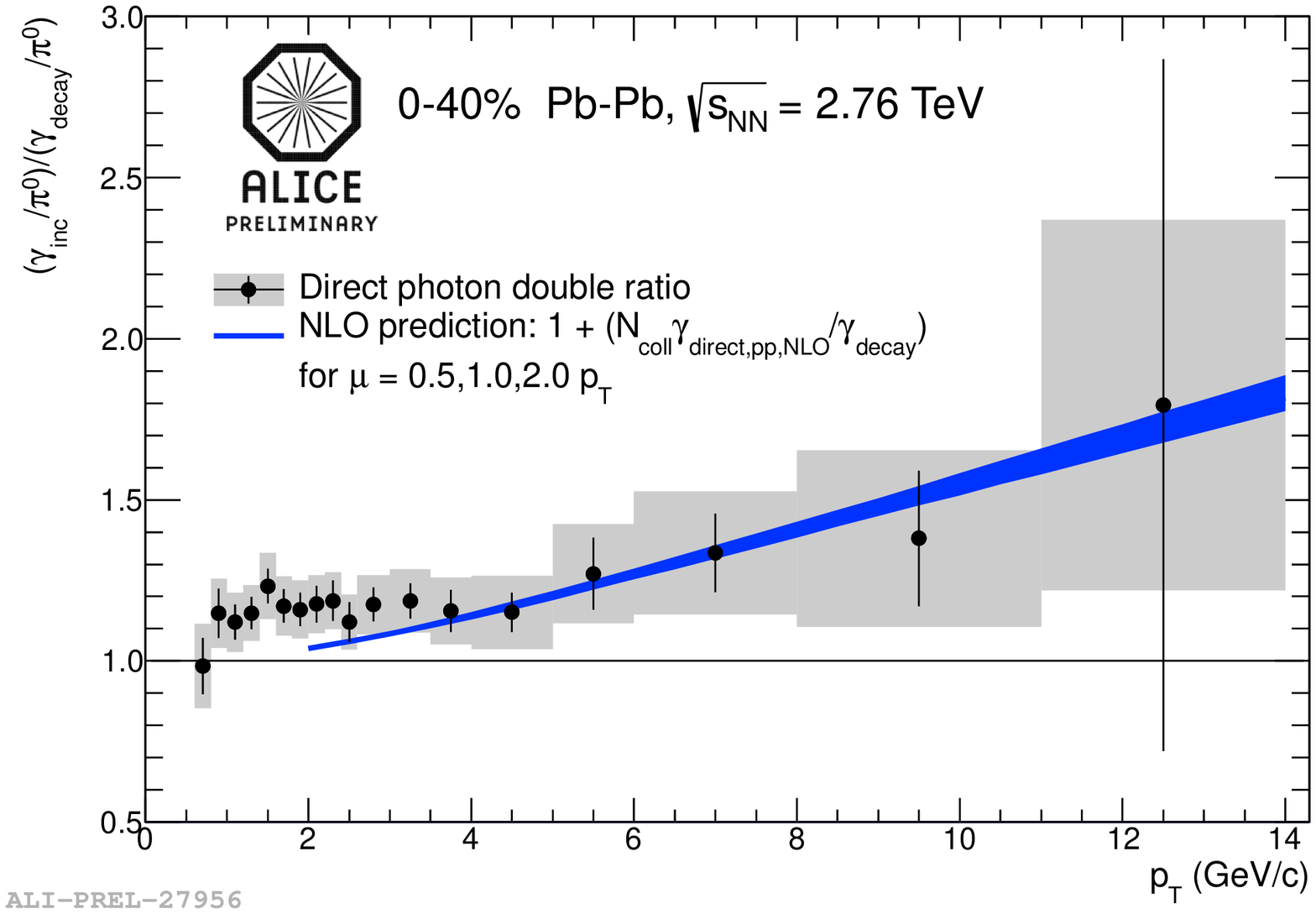}   
\end{subfigure}%
\begin{subfigure}{.5\textwidth}
  \centering
  \includegraphics[width=\textwidth]
  {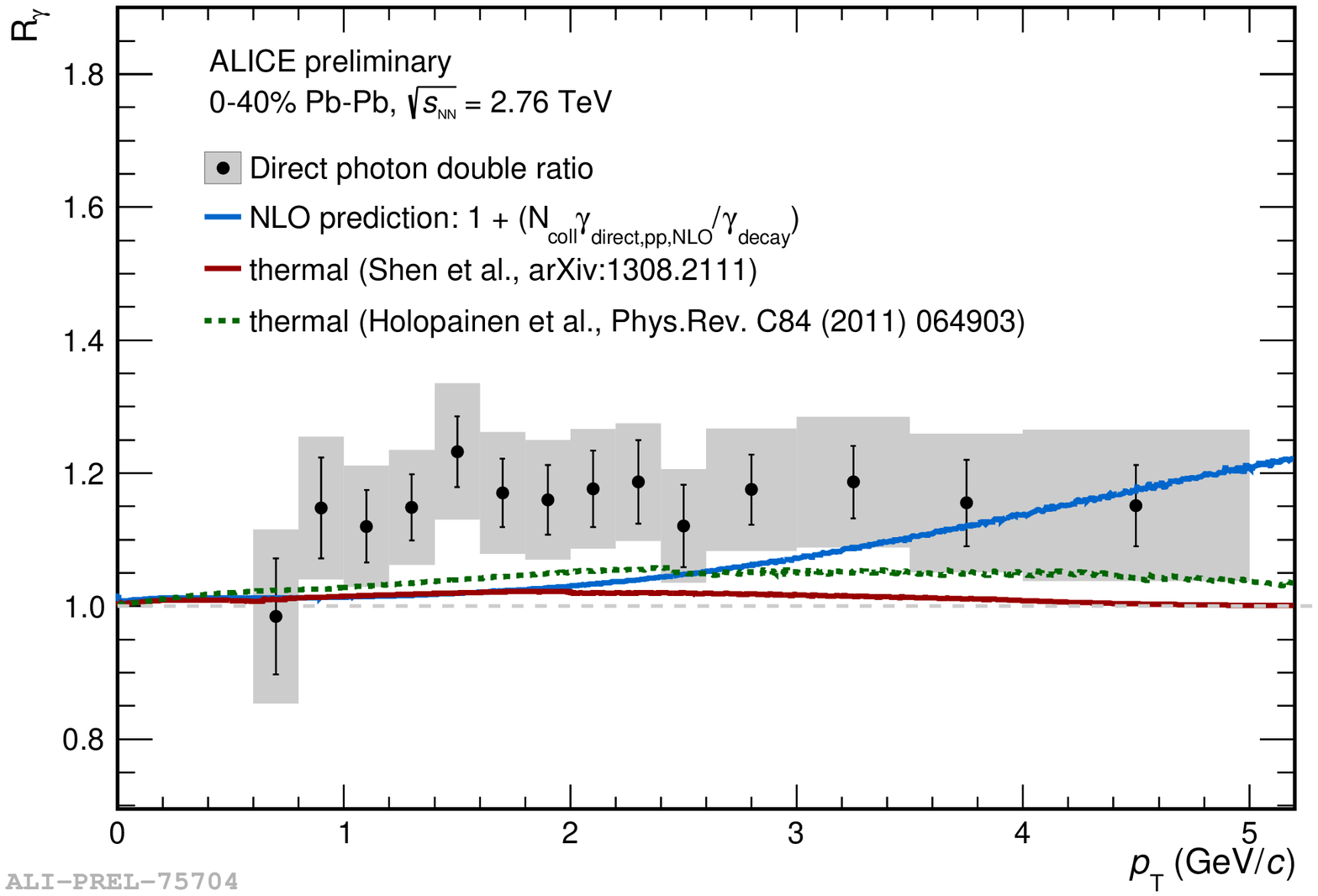}
\end{subfigure}
\caption{
Direct photon double ratio with NLO pQCD 
   at centrality 0-40\% 
  in \PbPb \mbox{} collisions at $\sqrt{s_{NN}}=2.76$ TeV \cite{Wilde}  (left).
  Measured double ratio
  with that predicted by hydrodynamic models (right).
}
\label{fig:DoubleR}
\end{figure}

~
Direct photons are defined as those
which do not originate from decay of unstable particles:

\begin{equation}
\label{eq:direct}
\mathrm{\gamma}_\mathrm{direct}=\mathrm{\gamma}_\mathrm{inc} - 
		\mathrm{\gamma}_\mathrm{decay}
		= \mathrm{\gamma}_\mathrm{inc} \cdot \left( 1 - \frac{\gamma_\mathrm{decay}}{\gamma_\mathrm{inc}} \right) 
		=\gamma_\mathrm{inc} \cdot \left(1-\frac{1}{R}\right),
\end{equation}
where $\gamma_{inc}$ and $\gamma_{decay}$ denote inclusive and decay
photon spectra, respectively, their ratio is
$R=\gamma_{inc}/\gamma_{decay}$. 
$R>$1 signals
the presence of direct photons.
In our analysis, we used a double ratio:

\begin{equation}
R_\mathrm{\gamma}=\frac{\gamma_\mathrm{inc}/\piz_\mathrm{meas}}
		{\gamma_\mathrm{decay}/\piz_\mathrm{param}}\approx R,
\end{equation}
where $\piz_\mathrm{meas}$ is the measured \piz\ spectrum
and $\piz_\mathrm{param}$ is its fitting function.
The advantage of this method is that major
systematic
uncertainties, associated with photon and \piz\ measurement,
cancel in the ratio.
The measured double ratio for \pp\ and \PbPb\ collisions is plotted
in Figures \ref{fig:Double_pp} and \ref{fig:DoubleR}
along with the Next to Leading Order (NLO) perturbative QCD
predictions.
$R_\mathrm{\gamma}$ for \pp \mbox{} collisions
is found to be in agreement with the NLO pQCD calculations
(left hand side
of Figure \ref{fig:Double_pp}).
The same agreement is observed for double ratio
in peripheral
(40-80\%) \PbPb
\mbox{} collisions (right hand side of Figure \ref{fig:Double_pp}),
whereas the result for central (0-40\%)
\PbPb \mbox{}
exhibits significant excess above 
the pQCD prediction at $\pt<$4 GeV/$c$,
as shown in Figure \ref{fig:DoubleR}.

\begin{figure}[H]
\begin{center}
\includegraphics[scale=0.5]
{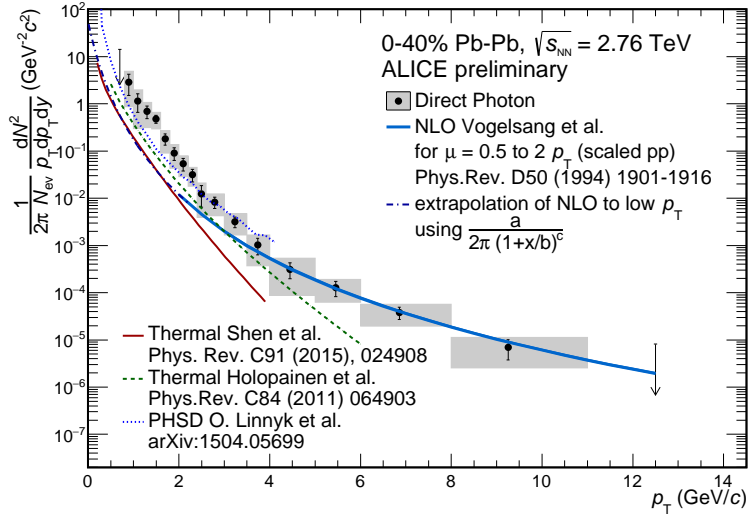}
\end{center}
\caption{\label{fig:DirectGamma}Direct photon spectrum at  
0-40\% in \PbPb\ at 2.76TeV with NLO
pQCD calculation and hydrodynamic model predictions.}
\end{figure}
The right hand side of
Figure \ref{fig:DoubleR} depicts the excess at $\pt \leq 5$ $\GeVc$ 
in more detail.
The measured double ratio exceeds that predicted by
hydrodynamic models.
The invariant yield of direct photons is plotted in 
Figure \ref{fig:DirectGamma}.
An exponential fit of the direct photon spectrum
of the form
$A \cdot \exp{(- \pt/T_{\mathrm{eff}})}$ for
$\pt \leq$2.2 GeV/$c$ yields an estimate for the inverse slope of the
exponent:
\begin{equation*}
T_{\mathrm{eff}} =304 \pm 51^{stat + sys}\mbox{  }\textrm{MeV}.
\end{equation*}

The obtained value of $T_\mathrm{eff}$ is larger
than the result obtained by the PHENIX experiment,
$T^\mathrm{RHIC}_\mathrm{eff}=219 \pm 19^{stat} \pm 19^{sys}$ $\mathrm{MeV}$
at $\sqrt{s_{NN}}$ =200 GeV \cite{PHENIX2}.
$T_\mathrm{eff}$ reflects the average temperature of
matter 
during its evolution but is also affected by
flow \cite{Shen}.
A consistent interpretation of $T_\mathrm{eff}$ in terms of the temperature
requires the understanding of the evolution of the fireball.
	\begin{wrapfigure}{R}{0.45 \textwidth}
				\vspace{-5pt}
				\begin{center}
				\includegraphics[width=50 mm]{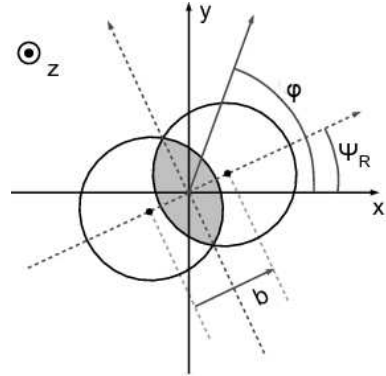}
			\end{center}
			\caption{\label{fig:coordinate} 
			Schematic view of a non-central ion-ion collision
			with non-zero impact parameter ${b}$.
			The direction of the line connecting
			the centres of the colliding nuclei 
			defines the orientation
			of the reaction plane expressed by the angle $\varPsi_R$
			used in (\ref{eq:flow}).
			}
			\end{wrapfigure}

The initial spatial anisotropy of non-central
ion-ion collision
leads to the build-up of collective anisotropic flow,
which is described in the distribution of 
final state particles as a function of azimuthal angle $\varphi$:
 \begin{equation}
 \frac{\mathrm{d}N}{\mathrm{d}\varphi} = \frac{N}{2\pi} \cdot \left(1+ 
		\mathrm{\sum_{n>1}}
		 v_\mathrm{n} \cos\mbox{}(n[\varphi- \varPsi_\mathrm{R}])\right),
		 \label{eq:flow}
 \end{equation}
where $\varPsi_\mathrm{R}$ is 
the orientation
of the reaction plane
as shown in Figure \ref{fig:coordinate}.
The thermal photons are expected to be produced at
an early phase of the collision and measurement of
their flow will help us to understand
development of the collective expansion 
at early times. 
Experimentally, one can measure the flow of inclusive photons,
estimate the flow of decay photons and
calculate the direct photon flow from the following expression:

\begin{equation}
\label{eq:flow2}
v_\mathrm{n}^{direct} = \frac{R_\mathrm{\gamma} \cdot v_\mathrm{n}^{inc} - 
					v_\mathrm{n} ^{decay} }
				{R_\mathrm{\gamma}-1}.
\end{equation}

\begin{figure}[H]
\centering
\begin{minipage}{.5\textwidth}
  \centering
  \includegraphics[width= \textwidth]{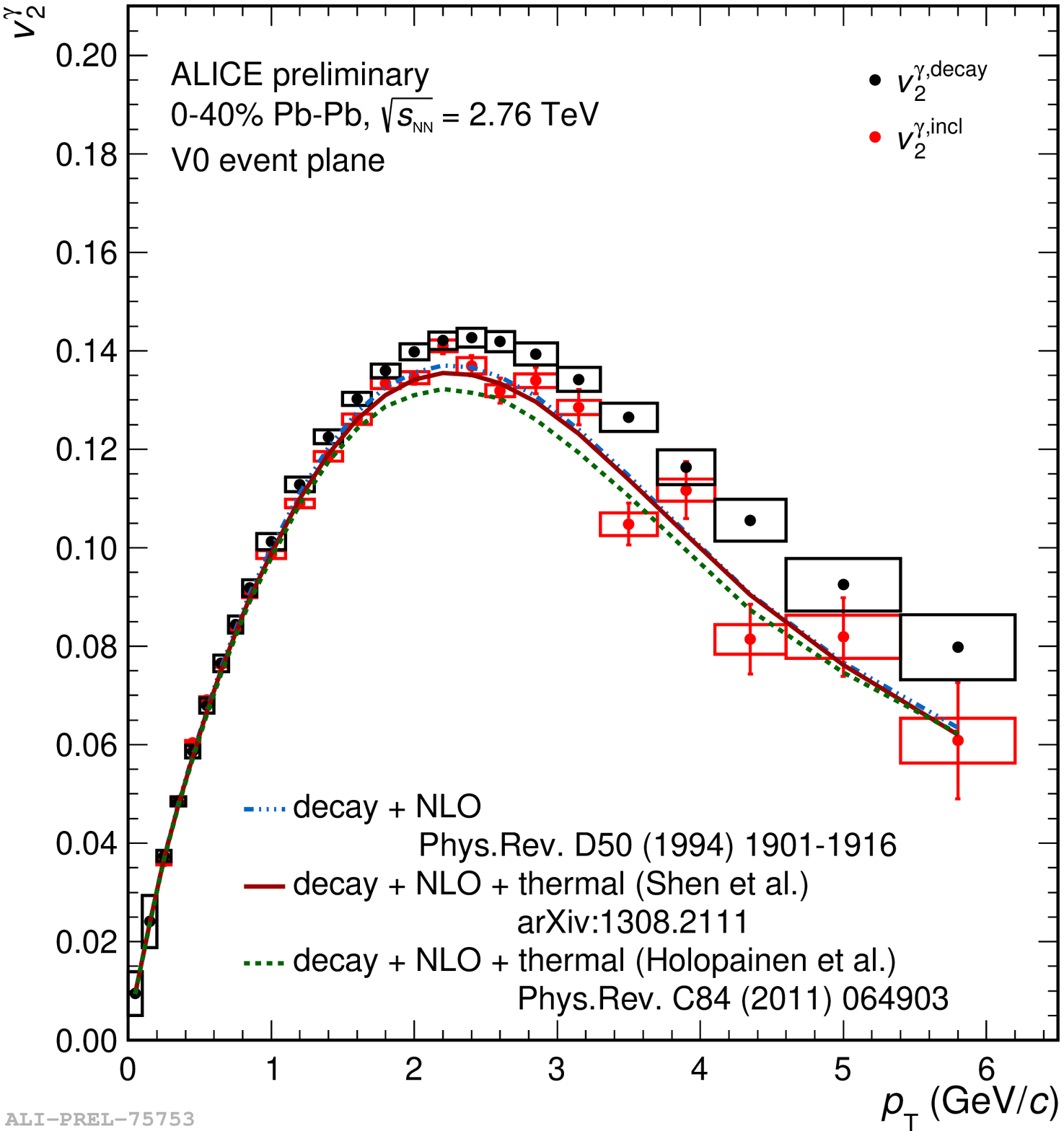}
    \captionof{figure}{Comparison of  elliptic flow
    of inclusive and decay \\ photons
   in 0-40\% \PbPb\ collisions at 2.76 TeV with theory.}
  \label{fig:v2Inc}
\end{minipage}%
\begin{minipage}{.5\textwidth}
  \centering
  \includegraphics[width= \textwidth]{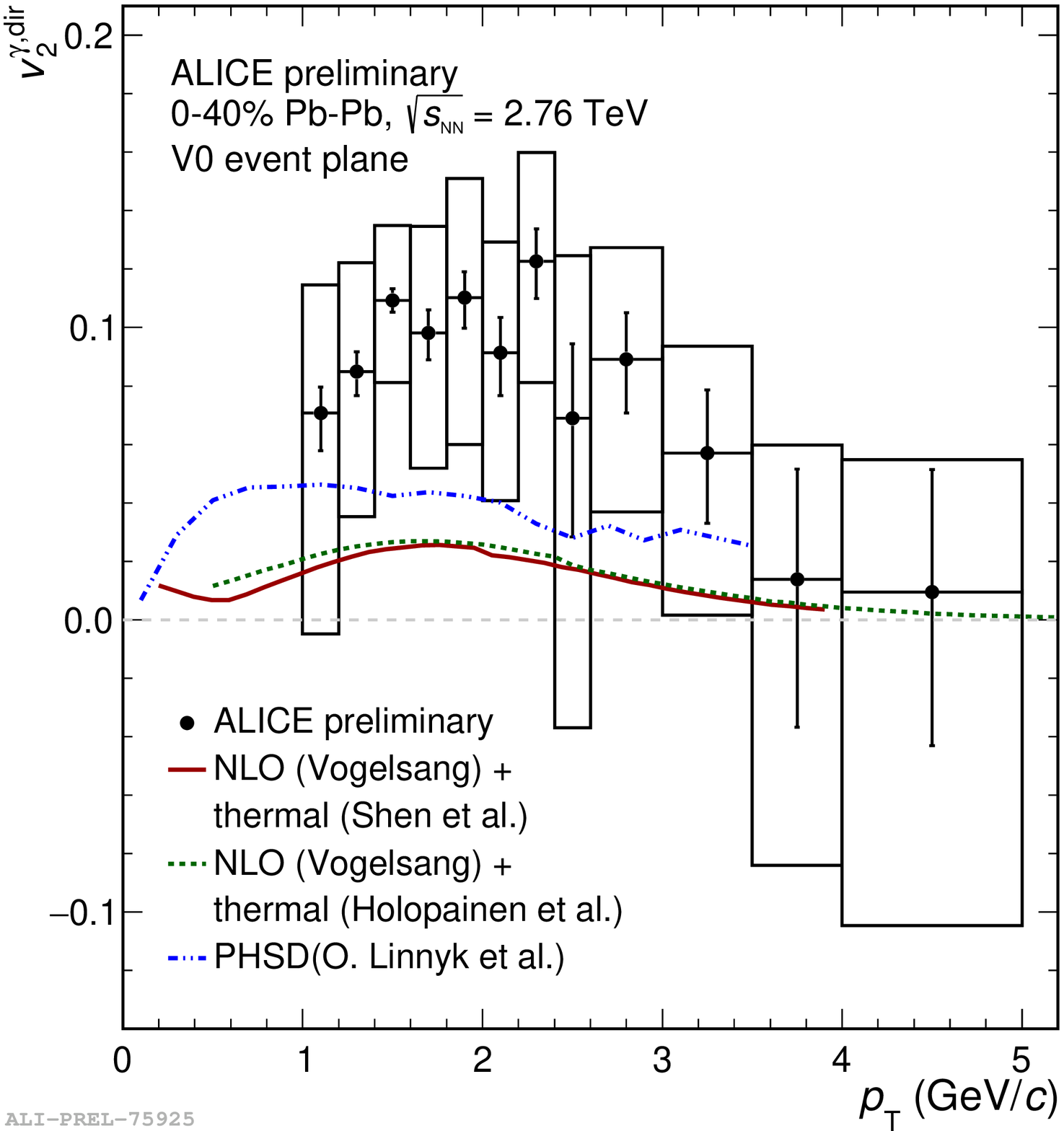}
  \captionof{figure}{Elliptic flow of direct photon measured in
  experiment with theory.}
  \label{fig:v2Dir}
\end{minipage}
\end{figure}

Figure \ref{fig:v2Inc} represents the $v_2$ coefficient
measuring
elliptic flow for inclusive and decay photons.
The elliptic flow of direct photons presented in
Figure \ref{fig:v2Dir} is non-zero in the same \pt \mbox{}
interval
where the excess of the yield over the perturbative QCD prediction
was observed.

\section{Summary}

~
We have presented the results of direct photon and \piz \mbox{}
production obtained from
proton-proton and lead-lead collisions in the ALICE experiment.
Measurement of neutral pions in pp collisions allowed to test
the validity of QCD-inspired models.
The QCD predictions agree with the measurements at
\sqrtS = 0.9 TeV but overpredict pion yields at
higher collision energies.
The nuclear modification factor for neutral pions
demonstrates a suppression of
the pion yield by factor 10 for the most central
collisions with respect to scaled pp collisions
at the same energy.
The study of jet-like, \piz\ -hadron
correlations reveals a strong away side suppression
and smaller near side
enhancement in Pb-Pb collisions due to in-medium effects.
 An excess of direct photons was found in \PbPb\ collisions
over that predicted
		from pQCD calculations and hydrodynamical models.
		The 
		effective temperature was found to be
		 $T_\mathrm{eff}=304 \pm 51^{stat + sys} \MeV$.
		A non-zero elliptic flow of direct photons
		was observed within the same \pt \mbox{}
		interval as the excess of thermal photons.


\begin{thebibliography}{99}

\bibitem{QGP} S. Borsanyi et al., JHEP {\bf 077}, 1011 (2010).

\bibitem{Martinez} G. Martinez, arXiv:1304.1452.

\bibitem{ALICE} K. Aamodt et al., J. of Instr.  {
\bf S08002}, 3 (2008).

\bibitem{pi0Paper} B. Abelev et al., Eur.Phys.J. {\bf C74}, 3108 (2014). 

\bibitem{pi0Paper2} B. Abelev et al.,  Phys.Lett. {\bf B717}, 162-172 (2012).

\bibitem{Florian} D. de Florian et al., arXiv:1410.6027 (2014).

\bibitem{PHENIX}  S. Adler et al., Phys.Rev.Lett. {\bf C76}, 034904 (2007).

\bibitem{Enhance} K. Aamodt et al., PRL {\bf 108}, 092301 (2012)

\bibitem{Wilde}  M. Wilde, Nucl.Phys. {\bf A904}, 573 (2013).

\bibitem{PHENIX2} A. Adare et al., Phys.Rev.Lett. {\bf 109}, 122302 (2012).

\bibitem{Shen} S. Chen, U. Heinz, Phys.Rev. {\bf C89},  4, 044910, (2014).



\end{thebibliography}
\end{document}